\documentclass[conference]{IEEEtran}
\IEEEoverridecommandlockouts
\usepackage{cite}
\usepackage{amsmath,amssymb,amsfonts}
\usepackage{algorithmic}
\usepackage{graphicx}
\usepackage{textcomp}
\usepackage{xcolor}
\usepackage{booktabs}
\usepackage{bm}
\usepackage{acronym}
\acrodef{6G}{sixth-generation}
\acrodef{EIT}{electromagnetic information theory}
\acrodef{MIMO}{multiple-input multiple-output}
\acrodef{CSI}{channel state information}
\acrodef{WF}{water filling}
\acrodef{SISO}{single-input single-output}
\acrodef{SNR}{signal-to-noise ratio}
\acrodef{VF}{vector fitting}
\acrodef{ULA}{uniform linear array}
\acrodef{SVD}{singular value decomposition}
\acrodef{NMSE}{normalized mean-square error}
\acrodef{AR}{autoregressive}
\acrodef{SA}{singularity-aware}
\acrodef{LR}{low rank}
\acrodef{TM}{transverse magnetic}
\acrodef{Tx}{transmitter}
\acrodef{Rx}{receiver}

\acrodef{MoM}{method of moments}
\AtBeginDocument{%
  \setlength{\abovedisplayskip}{1pt plus 0.5pt minus 0.5pt}%
  \setlength{\belowdisplayskip}{1pt plus 0.5pt minus 0.5pt}%
  \setlength{\abovedisplayshortskip}{0.5pt plus 0.5pt}%
  \setlength{\belowdisplayshortskip}{0.5pt plus 0.5pt}%
}

\makeatletter
\g@addto@macro\normalsize{\setlength{\baselineskip}{10.5pt}}
\makeatother
\def\BibTeX{{\rm B\kern-.05em{\sc i\kern-.025em b}\kern-.08em
    T\kern-.1667em\lower.7ex\hbox{E}\kern-.125emX}}

\title{Electromagnetic Singularity-Aware MIMO Channels: Scattering-Based Channel Modeling and Physics-Aware Power Allocation}

\author{
\IEEEauthorblockN{
Yubin Luo\IEEEauthorrefmark{1},
Jieao Zhu\IEEEauthorrefmark{2}\thanks{\textsuperscript{$\dagger$}Jieao Zhu was with the Department of Electronic Engineering,
Tsinghua University,  Haidian District, Beijing 100084,
China.},
Takumi Takahashi\IEEEauthorrefmark{1},
Hideki Ochiai\IEEEauthorrefmark{1}
}
\IEEEauthorblockA{
\IEEEauthorrefmark{1}
Graduate School of Engineering, The University of Osaka, Osaka, Japan\\
Email: \{y-luo, takahashi, ochiai\}@wcs.comm.eng.osaka-u.ac.jp;\quad
\IEEEauthorrefmark{2}zja17@tsinghua.org.cn
}

}

\begin{document}
\maketitle
\thispagestyle{empty}
\pagestyle{empty}
\begin{abstract}
\Ac{EIT} brings electromagnetic physics into wireless channel analysis and offers a physical view of channel structures hidden by conventional models.
This paper studies channel instability near electromagnetic singular states, where small environmental perturbations can cause large variations in the end-to-end channel.
Using resolvent-based spectral analysis, we develop an electromagnetic \ac{MIMO} channel model that decomposes the end-to-end channel into a smooth regular component and singularity-associated components.
The analysis shows that the singularity-associated components become more sensitive to environmental perturbations near singular states.
The proposed model also reveals a limitation of conventional \ac{WF}: power allocation based only on instantaneous channel gains does not
account for such instability.
We therefore estimate singularity information from broadband \ac{MIMO} \ac{CSI} and develop a singularity-aware transmit covariance optimization method.
Simulations demonstrate numerical stability of the estimated pole
frequencies and illustrate the environmental sensitivity of
pole-associated channel components.
The proposed allocation reduces weighted singular-channel exposure by $90.32\%$ while retaining $95\%$ of the nominal \ac{WF} rate.
\end{abstract}

\acresetall
\section{Introduction}

Recent studies of \ac{6G} wireless channels have emphasized the need for more physical channel descriptions~\cite{zhang2020channel}.
Within this context, \ac{EIT} has emerged as an important research direction~\cite{zhu2025eit}.
\ac{EIT} aims to integrate electrodynamic descriptions based on Maxwell's equations with conventional wireless communication theories, including information theory, multi-antenna communications, and signal processing, so as to reinterpret wireless channels and communication systems from their underlying electromagnetic physics.
Unlike conventional channel models that typically represent propagation through statistical, parametric, or geometry-based descriptions, \ac{EIT} starts from the intrinsic physical properties of electromagnetic fields and focuses on how the underlying wave phenomena govern channel structure and, in turn, affect transmission performance~\cite{zhu2026capacity,yuan2025normalization}.

Within this framework, physics-based channel modeling is an important component of \ac{EIT}.
By explicitly incorporating electromagnetic propagation and scattering mechanisms, such models can reveal channel structures that are often hidden in conventional statistical representations.
Recent works have demonstrated that this perspective can provide new insights into
channel characterization and estimation~\cite{wei2025hmimo,pizzo2020holographic,jiang2024sensing}.
Our previous study analyzed the ill-conditioning of \ac{CSI}-based inverse scattering for material reconstruction and the joint roles of internal scattering and \ac{Tx}/\ac{Rx} propagation operators~\cite{luo2026illposed}.
These results motivate a more explicit investigation of scattering-induced channel structures for communication modeling and design.

Motivated by the above observations, we develop a new \ac{MIMO} channel model from the perspective of electromagnetic scattering to explicitly characterize channel instability induced by underlying electromagnetic singular structures.
Unlike conventional models that absorb scattering effects into random coefficients or statistical parameters, the proposed model is derived through resolvent-based spectral analysis and provides a structured decomposition of the end-to-end channel, establishing an explicit connection between such instability and the scattering environment, system coupling, and operating frequency.
We further analyze this behavior from the perspectives of the eigenvalue spectrum and electromagnetic modes, showing that, as the channel approaches an electromagnetic singular state, the associated spectral components and communication modes become increasingly sensitive to environmental perturbations.
Furthermore, the proposed channel model reveals a limitation of conventional \ac{WF}~\cite{telatar1999capacity}.
Since \ac{WF} allocates power according to instantaneous channel gains, it does not account for the environmental sensitivity associated with singular channel components and may therefore assign substantial power to unstable channel directions.
Channel aging has likewise motivated robust \ac{MIMO} precoding under imperfect \ac{CSI}~\cite{wang2024aging}.
Motivated by this observation, we develop a singularity-aware power allocation scheme that estimates the relevant singularity information from \ac{CSI} and incorporates it into transmit covariance optimization, thereby suppressing power allocation to unstable channel components while promoting the utilization of regular channel components.

The remainder of this paper is organized as follows.
Section~II develops the electromagnetic \ac{MIMO} channel model and its singularity characterization.
Section~III presents the singular-parameter estimation method and the singularity-aware power allocation scheme.
Section~IV provides simulation results, and Section~V concludes the paper.

\section{Electromagnetic Channel Model}

In this section, we first establish the scattering-based channel model and then analyze its instability and singularity in terms of the underlying electromagnetic propagation mechanism.

\subsection{End-to-End Electromagnetic \ac{MIMO} Channel}

Consider a two-dimensional transverse magnetic-to-$z$ (\acs{TM}$_z$)\acused{TM} electromagnetic environment with an out-of-plane electric field.
Let $\mathcal D\subset\mathbb R^2$ denote a bounded inhomogeneous scattering region, with field space $\mathcal H_{\mathcal D}\triangleq L^2(\mathcal D)$.
The system consists of $N_t$ transmit antennas and $N_r$ receive antennas.
Wideband measurements are collected at $K$ frequencies $f_k$ in the measurement band, which need not be uniformly spaced.
Define
\begin{equation}
\omega_k=2\pi f_k,\qquad z_k=\omega_k,\qquad
k=1,\ldots,K.
\end{equation}
Here $\omega$ denotes real angular frequency; $z\in\mathbb C$ is its analytic continuation, so $z=\omega$ on the measurement axis.
We set $z_c=2\pi f_c$ for the band-center frequency $f_c$ and use $\mathrm j^2=-1$ throughout.
The subscript in \ac{TM}$_z$ denotes the out-of-plane spatial direction, independently of complex frequency $z$.

The material distribution in $\mathcal D$ is described by the complex contrast
\begin{equation}
\chi(\mathbf r;\omega)
=
\varepsilon_r(\mathbf r)-1
+
\mathrm j
\frac{\sigma_e(\mathbf r)}
{\varepsilon_0\omega},
\qquad
\mathbf r\in\mathcal D,
\label{eq:contrast}
\end{equation}
where $\varepsilon_r$, $\sigma_e$, and $\varepsilon_0$ are relative permittivity, conductivity, and free-space permittivity.
The material response is assumed to be weakly dispersive over the considered bandwidth; $\chi(\mathbf r;z)$ denotes its analytic continuation to complex angular frequency.

The incident field $E^i$ and total field $E^t$ satisfy the Lippmann--Schwinger equation~\cite{jiang2024sensing,chew1995waves}
\begin{equation}
E^t(\mathbf r,\omega)
=
E^i(\mathbf r,\omega)
+
\int_{\mathcal D}
G(\mathbf r,\mathbf r';\omega)
\chi(\mathbf r';\omega)
E^t(\mathbf r',\omega)
\,\mathrm d\mathbf r',
\label{eq:ls_equation}
\end{equation}
where
$
G(\mathbf r,\mathbf r';\omega)
=
-\frac{\mathrm j\kappa^2}{4}
H_0^{(2)}
\left(
\kappa\|\mathbf r-\mathbf r'\|_2
\right),
$
with $\kappa=\omega/c_0$, and $c_0$ is the free-space wave velocity.
$H_0^{(2)}(\cdot)$ is the zeroth-order Hankel function of the second kind.
For vectors, $\|\cdot\|_2$ denotes the Euclidean norm.

To facilitate the spectral analysis of the frequency-dependent scattering operator, the real angular frequency is analytically continued to a complex
variable $z$.
On $\mathcal H_{\mathcal D}$, define the Green operator by $[\mathcal G_D(z)u](\mathbf r)=\int_{\mathcal D}G(\mathbf r,\mathbf r';z)u(\mathbf r')\,\mathrm d\mathbf r'$ and the contrast multiplication operator by
$[\mathcal X_\chi(z)u](\mathbf r)=\chi(\mathbf r;z)u(\mathbf r)$.
With $\mathcal I$ the identity on $\mathcal H_{\mathcal D}$, set
\begin{equation}
\mathcal K_\chi=\mathcal G_D\mathcal X_\chi,\quad
\widetilde{\mathcal R}_\chi=(\mathcal I-\mathcal K_\chi)^{-1},\quad
\mathcal R_\chi=\mathcal X_\chi\widetilde{\mathcal R}_\chi.
\label{eq:scattering_resolvent}
\end{equation}
All operators depend on $z$.
The Lippmann--Schwinger equation is $(\mathcal I-\mathcal K_\chi)E^t=E^i$; the scattering-response operator $\mathcal R_\chi$ maps the incident field to the equivalent source $\mathcal X_\chi E^t$.

Let $\mathbf B(z):\mathbb C^{N_t}\to\mathcal H_{\mathcal D}$ denote the transmit-to-domain free-space Green operator that maps transmit signals to incident fields in $\mathcal D$, and let $\mathbf C(z):\mathcal H_{\mathcal D}\to\mathbb C^{N_r}$ denote the domain-to-receiver free-space Green operator that maps equivalent scattering sources to receive signals.
The scattered end-to-end channel is
\begin{equation}
\mathbf H_{\rm sc}(z)
=
\mathbf C(z)\mathcal R_\chi(z)\mathbf B(z)
\in\mathbb C^{N_r\times N_t}.
\label{eq:end_to_end_channel}
\end{equation}
The total channel is $\mathbf H=\mathbf H_{\rm dir}+\mathbf H_{\rm sc}$, where $\mathbf H_{\rm dir}$ is the known free-space direct term, absorbed into the analytic regular component and subtracted before pole fitting.
For communication, $\mathbf H(z)$ denotes the sum of the direct and scattered components.
At $z_k=\omega_k$, write
$\mathbf H_k=\mathbf H(z_k)$ and
\begin{equation}
\mathbf y_k
=
\mathbf H_k\mathbf x_k+\mathbf n_k,\qquad
\mathbf n_k\sim\mathcal{CN}
(\mathbf0,N_0\mathbf I_{N_r}).
\label{eq:frequency_mimo}
\end{equation}
Here $\mathbf x_k\in\mathbb C^{N_t}$ and $\mathbf y_k\in\mathbb C^{N_r}$ are the transmit and receive vectors; $\mathcal{CN}$ denotes a circular complex Gaussian distribution, and $N_0$ is the noise power per receive dimension.

\subsection{Regular--Singular Channel Decomposition}

To characterize the singular structure of the scattering response, we consider the spectrum of the compact operator $\mathcal K_\chi(z)$.
The resolvent $(\mathcal I-\mathcal K_\chi(z))^{-1}$ becomes singular when $\mathcal I-\mathcal K_\chi(z)$ is noninvertible, equivalently when an eigenvalue of $\mathcal K_\chi(z)$ reaches unity.
We refer to the corresponding complex frequency $p_j$ as an electromagnetic pole.
Following analytic perturbation theory~\cite{kato1995perturbation}, let $\lambda_j(z)$ be an isolated simple eigenvalue branch satisfying
\begin{equation}
\lambda_j(p_j)=1,\qquad
\lambda_j'(p_j)
=
\left.\frac{\mathrm d\lambda_j(z)}{\mathrm dz}\right|_{z=p_j}
\neq0,
\end{equation}
where the prime denotes differentiation with respect to the complex frequency variable $z$.
Its Riesz projection is
\begin{equation}
\Pi_j(z)
=
\frac{1}{2\pi\mathrm j}
\oint_{\Gamma_j}
[\zeta\mathcal I-\mathcal K_\chi(z)]^{-1}
\,\mathrm d\zeta,
\label{eq:riesz_projection}
\end{equation}
where $\Gamma_j$ encloses only $\lambda_j(z)$.
The projection $\Pi_j(z)$ isolates the invariant spectral subspace associated with $\lambda_j(z)$ from the remaining spectrum.
On the corresponding invariant subspace, the resolvent $\widetilde{\mathcal R}_\chi(z)
\triangleq[\mathcal I-\mathcal K_\chi(z)]^{-1}$ satisfies
\[
\widetilde{\mathcal R}_\chi(z)\Pi_j(z)
=
\frac{\Pi_j(z)}{1-\lambda_j(z)},
\]
while the contribution from the remaining spectrum is analytic near
$p_j$.
Using
$1-\lambda_j(z)
=-\lambda_j'(p_j)(z-p_j)+\mathcal O((z-p_j)^2)$
and
$\Pi_j(z)=\Pi_j(p_j)+\mathcal O(z-p_j)$,
the resolvent admits the local Laurent expansion
\begin{equation}
\widetilde{\mathcal R}_\chi(z)
=
-\frac{\Pi_j(p_j)}
{\lambda_j'(p_j)(z-p_j)}
+
\widetilde{\mathcal R}_{{\rm reg},j}(z),
\label{eq:resolvent_laurent}
\end{equation}
where $\widetilde{\mathcal R}_{{\rm reg},j}(z)$ is analytic at $p_j$.

Substituting \eqref{eq:resolvent_laurent} into the end-to-end channel
model yields
\begin{equation}
\mathbf H(z)
=
\mathbf H_{\rm reg}(z)
+
\sum_{j=1}^{J}
\frac{\mathbf R_j}{z-p_j},
\label{eq:regular_singular_channel}
\end{equation}
where
\begin{equation}
\mathbf R_j
=
-\frac{
\mathbf C(p_j)\mathcal X_\chi(p_j)
\Pi_j(p_j)\mathbf B(p_j)
}{
\lambda_j'(p_j)
}
\label{eq:end_to_end_residue}
\end{equation}
is the end-to-end residue associated with the $j$th electromagnetic
pole, and $\mathbf H_{\rm reg}(z)$ collects the direct channel and all
terms analytic in the considered region. Equation
\eqref{eq:regular_singular_channel} therefore gives the
regular--singular decomposition of the \ac{MIMO} channel induced by the
electromagnetic scattering spectrum.

The pole-associated terms are particularly sensitive near an
electromagnetic singular state. At the operator level, the corresponding
modal response is proportional to $(1-\lambda_j)^{-1}$, whose first-order
variation satisfies
$\delta(1-\lambda_j)^{-1}\simeq\delta\lambda_j/(1-\lambda_j)^2$.
Hence, as $\lambda_j(z)$ approaches the singular condition
$\lambda_j(z)=1$, small perturbations of the electromagnetic spectrum can
produce large variations in the associated modal response.

The same behavior appears directly in the channel representation. Let a
small environmental perturbation induce changes
$\delta p_j$ and $\delta\mathbf R_j$. To first order,
\begin{equation}
\delta
\left(
\frac{\mathbf R_j}{z-p_j}
\right)
\simeq
\frac{\delta\mathbf R_j}{z-p_j}
+
\frac{\mathbf R_j\,\delta p_j}
{(z-p_j)^2}.
\label{eq:pole_channel_perturbation}
\end{equation}
Thus, the variation caused by pole displacement grows as
$|z-p_j|^{-2}$. When the operating frequency approaches an
electromagnetic pole, even a small environment-induced spectral
perturbation can therefore cause a pronounced channel variation. In this
sense, the corresponding channel component is intrinsically unstable with
respect to environmental perturbations.

Meanwhile, its nominal magnitude satisfies
$\|\mathbf R_j/(z-p_j)\|_2=\|\mathbf R_j\|_2/|z-p_j|$,
so the same component may also strongly affect the instantaneous channel
gain. Note that $\lambda_j(z)$ is an eigenvalue of the electromagnetic
scattering operator, whereas the singular values of the end-to-end \ac{MIMO}
channel characterize communication gains; the former influences the
latter through the residue and the transmit/receive coupling.

At the $k$th complex-frequency sample $z_k$, let
$\mathbf H_k=\mathbf U_k\boldsymbol\Sigma_k\mathbf V_k^H$,
where $\sigma_{k,i}$ and $\mathbf v_{k,i}$ are the $i$th singular value
and right-singular vector at frequency $k$, respectively, and
$N_{s,k}=\operatorname{rank}(\mathbf H_k)$.
The above analysis shows that different channel directions can exhibit different levels of electromagnetic sensitivity, which directly affects how transmit power should be distributed across spatial--frequency modes.
As a reference, conventional \ac{WF} over all frequencies and
spatial modes assigns
\begin{equation}
p_{k,i}^{\text{\acs{WF}}}
=
\left[
\mu-\frac{N_0}{\sigma_{k,i}^2}
\right]^+,
\qquad
\sum_{k=1}^{K}
\sum_{i=1}^{N_{s,k}}
p_{k,i}^{\text{\acs{WF}}}
=
P,
\label{eq:water_filling}
\end{equation}
where $[a]^+=\max(a,0)$, $\mu$ is the water level, and $P$ is the total power budget.
Water-filling is optimal for the nominal instantaneous channel under the conventional rate-maximization objective, but it depends
only on $\{\sigma_{k,i}\}$ and does not account for the underlying
electromagnetic sensitivity.
Consequently, a channel direction can receive substantial transmit power even when its gain is strongly associated with
an environment-sensitive singular component.
This motivates the singularity-aware transmission design developed in the following section.

\section{Singularity-Aware Channel Estimation and Transmission Design}

\subsection{Singular-Parameter Channel Estimation}

We estimate shared poles and link-dependent residues from broadband \ac{MIMO} \ac{CSI} and use the estimates for subsequent rate-constrained power allocation.

For a rank-one Riesz projection, the end-to-end residue matrix can be written as
\begin{equation}
\mathbf R_j(\mathbf a)
=
-\frac{
\mathbf c_j
\mathbf b_j^{H}(\mathbf a)
}{
\lambda_j'(p_j)
},
\label{eq:rank_one_residue}
\end{equation}
where $\mathbf c_j\in\mathbb C^{N_r}$ and
$\mathbf b_j(\mathbf a)\in\mathbb C^{N_t}$ are the receive- and
transmit-side coupling vectors, respectively. For right and left
eigenfunctions $u_j,\ell_j\in\mathcal H_{\mathcal D}$ normalized so that
$\Pi_j(p_j)=u_j\ell_j^H$ and $\ell_j^Hu_j=1$,
$\mathbf c_j=\mathbf C(p_j)\mathcal X_\chi(p_j)u_j$ and
$\mathbf b_j^H(\mathbf a)=\ell_j^H\mathbf B(p_j;\mathbf a)$.
The geometry $\mathbf a$ collects the fixed transmit-element locations;
$\mathbf B(z;\mathbf a)$ is the propagation operator defined in Section~II.

Thus, mode $j$ is physically characterized by
$(p_j,\lambda_j'(p_j),\mathbf c_j,\mathbf b_j(\mathbf a))$.
Since the scaling among
$\lambda_j'(p_j)$, $\mathbf c_j$, and $\mathbf b_j(\mathbf a)$
is not uniquely identifiable from end-to-end \ac{CSI}, we first estimate
$\{p_j,\mathbf R_j\}$ and then obtain an equivalent four-parameter
representation from the \ac{SVD} of $\mathbf R_j$.

Let $g_{mn}(z)$ denote the $(m,n)$th entry of 
$\mathbf H_{\rm sc}(z)=\mathbf H(z)-\mathbf H_{\rm dir}(z)$, with
$m=1,\ldots,N_r$ and $n=1,\ldots,N_t$. From \eqref{eq:regular_singular_channel}, the regular component
$\mathbf H_{\rm reg}(z)-\mathbf H_{\rm dir}(z)$ is analytic and can be locally approximated
by its first-order Taylor expansion around the center frequency $z_c$.
Taking the $(m,n)$th entry gives
\begin{equation}
g_{mn}(z_k)
\approx
d_{mn}+e_{mn}z_k
+
\sum_{j=1}^{J}
\frac{r_{j,mn}}{z_k-p_j},
\label{eq:scalar_csi_model}
\end{equation}
where
$e_{mn}=g_{{\rm reg},mn}'(z_c)$,
$d_{mn}=g_{{\rm reg},mn}(z_c)-z_c e_{mn}$,
and $r_{j,mn}$ is the $(m,n)$th entry of $\mathbf R_j$.
The poles $\{p_j\}$ are shared across all \ac{MIMO} links, whereas the
residues are link dependent.

\subsubsection{Shared-Pole Vector Fitting}

To jointly estimate the shared poles $\{p_j\}_{j=1}^{J}$ from broadband
\ac{MIMO} \ac{CSI}, we employ shared-pole \ac{VF}~\cite{gustavsen1999vector,deschrijver2008multiport}. At iteration $\ell$,
a set of distinct temporary poles
$\{\xi_j^{(\ell)}\}_{j=1}^{J}$ is introduced, with monic polynomial
$\Xi^{(\ell)}(z)\triangleq\prod_{j=1}^{J}(z-\xi_j^{(\ell)})$.
The temporary poles define the rational basis used in the current
iteration and are not assumed to coincide with the physical poles.

Under the first-order approximation of the regular channel background, \eqref{eq:scalar_csi_model} can be written as $g_{mn}(z)\simeq n_{mn}(z)/q(z)$, where
$
q(z)
\triangleq
\prod_{j=1}^{J}(z-p_j),
$
and
\begin{equation}
n_{mn}(z)
\triangleq
\left(d_{mn}+e_{mn}z\right)q(z)
+
\sum_{j=1}^{J}
r_{j,mn}
\frac{q(z)}{z-p_j}.
\end{equation}
Here, $q(z)$ is common to all \ac{MIMO} links, whereas $n_{mn}(z)$ is
link dependent.

The key step of \ac{VF} is to reparameterize the same rational response with
respect to the temporary denominator $\Xi^{(\ell)}(z)$:
\begin{equation}
g_{mn}(z)
\simeq
\frac{
n_{mn}(z)/\Xi^{(\ell)}(z)
}{
q(z)/\Xi^{(\ell)}(z)
}.
\label{eq:vf_reparameterization}
\end{equation}
Since both $q(z)$ and $\Xi^{(\ell)}(z)$ are monic polynomials of degree $J$,
their ratio approaches unity as $|z|\rightarrow\infty$, and its poles are
located at $\{\xi_j^{(\ell)}\}$. Therefore,
\begin{equation}
\frac{q(z)}{\Xi^{(\ell)}(z)}
=
1+
\sum_{j=1}^{J}
\frac{\upsilon_j^{(\ell)}}
{z-\xi_j^{(\ell)}}
\triangleq
\eta^{(\ell)}(z).
\label{eq:vf_denominator}
\end{equation}
The coefficients $\{\upsilon_j^{(\ell)}\}$ are shared across all \ac{MIMO} links
because they represent the common physical denominator.

For the numerator, $\deg n_{mn}\leq J+1$, whereas
$\deg \Xi^{(\ell)}=J$. Hence, polynomial division yields a first-order
polynomial plus a proper rational remainder, which can be expanded over
the same temporary poles:
\begin{equation}
\frac{n_{mn}(z)}{\Xi^{(\ell)}(z)}
=
A_{mn}^{(\ell)}
+
B_{mn}^{(\ell)}z
+
\sum_{j=1}^{J}
\frac{\widetilde r_{j,mn}^{(\ell)}}
{z-\xi_j^{(\ell)}}.
\label{eq:vf_numerator_expansion}
\end{equation}
The coefficients
$A_{mn}^{(\ell)}$, $B_{mn}^{(\ell)}$, and
$\widetilde r_{j,mn}^{(\ell)}$
are link dependent and vary with the current \ac{VF} iteration.

Substituting \eqref{eq:vf_denominator} and
\eqref{eq:vf_numerator_expansion} into
\eqref{eq:vf_reparameterization} gives the barycentric representation
\begin{equation}
g_{mn}(z)
\simeq
\frac{
A_{mn}^{(\ell)}
+
B_{mn}^{(\ell)}z
+
\displaystyle\sum_{j=1}^{J}
\frac{\widetilde r_{j,mn}^{(\ell)}}
{z-\xi_j^{(\ell)}}
}{
1+
\displaystyle\sum_{j=1}^{J}
\frac{\upsilon_j^{(\ell)}}
{z-\xi_j^{(\ell)}}
}.
\label{eq:vf_barycentric}
\end{equation}

For fixed $\{\xi_j^{(\ell)}\}$, the basis functions $(z-\xi_j^{(\ell)})^{-1}$ are known.
Cross-multiplying \eqref{eq:vf_barycentric} yields an equation that is linear in the unknown coefficients.
For the fitting sample $\hat g_{mn,k}$ at $z_k$, define
\begin{equation}
\begin{aligned}
\epsilon_{mn,k}^{(\ell)}
={}&
A_{mn}^{(\ell)}
+
B_{mn}^{(\ell)}z_k
-
\hat g_{mn,k}
\\
&+
\sum_{j=1}^{J}
\frac{
\widetilde r_{j,mn}^{(\ell)}
-
\hat g_{mn,k}\upsilon_j^{(\ell)}
}{
z_k-\xi_j^{(\ell)}
}.
\end{aligned}
\label{eq:vf_residual}
\end{equation}
The coefficients are jointly estimated by
\begin{equation}
\min_{
\substack{
\{A_{mn}^{(\ell)},B_{mn}^{(\ell)},
\widetilde r_{j,mn}^{(\ell)}\}\\
\{\upsilon_j^{(\ell)}\}
}}
\sum_{m=1}^{N_r}
\sum_{n=1}^{N_t}
\sum_{k=1}^{K}
\left|
\epsilon_{mn,k}^{(\ell)}
\right|^2.
\label{eq:shared_vf_ls}
\end{equation}
The numerator coefficients are link dependent, whereas
$\{\upsilon_j^{(\ell)}\}$ are common to all links. Hence, all
$N_rN_tK$ \ac{CSI} samples jointly constrain the same physical pole
structure.

Finally, since ideally
$\eta^{(\ell)}(z)=q(z)/\Xi^{(\ell)}(z)$,
the zeros of $\eta^{(\ell)}(z)$ correspond to the zeros of $q(z)$, i.e.,
the physical poles. They are therefore used to update the temporary
poles in the next iteration.

\subsubsection{Pole Relocation and Singular-Parameter Recovery}

Update $\{\xi_j^{(\ell+1)}\}_{j=1}^{J}$ to the $J$ zeros of
$\eta^{(\ell)}(z)$. These are the zeros of the monic polynomial
$\Xi^{(\ell)}(z)\eta^{(\ell)}(z)$ after cancellation of its apparent
temporary-pole singularities. The joint least-squares fitting and pole
relocation are repeated until convergence; denote the resulting shared
estimates by $\{\hat p_j\}$.

With $\{\hat p_j\}$ fixed, define the final-fit residual
\begin{equation}
\epsilon^{\rm fin}_{mn,k}
=
\hat g_{mn,k}
-
d_{mn}
-
e_{mn}z_k
-
\sum_{j=1}^{J}
\frac{r_{j,mn}}{z_k-\hat p_j}.
\label{eq:final_fit_residual}
\end{equation}
The final link-dependent coefficients minimize
\begin{equation}
\min_{\{d_{mn},e_{mn},r_{j,mn}\}}
\sum_{m=1}^{N_r}
\sum_{n=1}^{N_t}
\sum_{k=1}^{K}
|\epsilon^{\rm fin}_{mn,k}|^2.
\label{eq:final_residue_ls}
\end{equation}

Denote the minimizing coefficients by
$\hat d_{mn}$, $\hat e_{mn}$, and $\hat r_{j,mn}$.
The estimated residues are assembled into the end-to-end residue matrix
$\hat{\mathbf R}_j=[\hat r_{j,mn}]\in\mathbb C^{N_r\times N_t}$.
We then perform the \ac{SVD}
\begin{equation}
\hat{\mathbf R}_j
=
\hat{\mathbf U}_j
\hat{\boldsymbol\Sigma}_j
\hat{\mathbf V}_j^H
\approx
\hat\sigma_{j,1}
\hat{\mathbf u}_{j,1}
\hat{\mathbf v}_{j,1}^H,
\label{eq:residue_svd}
\end{equation}
where $\hat{\mathbf U}_j$ and $\hat{\mathbf V}_j$ contain the left and
right singular vectors, respectively,
$\hat{\boldsymbol\Sigma}_j$ contains the singular values, and the
approximation retains the leading singular triplet.

For transmission design, let $L_j$ be the retained residue rank and keep
the first $L_j$ singular triplets in
$\hat{\mathbf R}_j^{\text{\acs{LR}}}=\sum_{i=1}^{L_j}\hat\sigma_{j,i}
\hat{\mathbf u}_{j,i}\hat{\mathbf v}_{j,i}^H$. The leading triplet
provides a compact physical summary, while the retained matrix preserves
the residue energy used for power allocation.

The leading singular triplet yields the identifiable equivalent parameters
$(\hat p_j,\hat\rho_j,\hat{\mathbf c}_j,\hat{\mathbf b}_j)$,
where $\hat\rho_j=\hat\sigma_{j,1}$,
$\hat{\mathbf c}_j=\hat{\mathbf u}_{j,1}$, and
$\hat{\mathbf b}_j=\hat{\mathbf v}_{j,1}$. The unit-norm vectors fix their
scale, while a common phase is immaterial to their outer product.
These equivalent parameters retain the pole location, residue magnitude,
and the transmit- and receive-side spatial coupling directions without
claiming a separate estimate of the internal eigenvalue slope.

\subsection{Singularity-Aware Power Allocation}

Using the low-rank residues $\hat{\mathbf R}_j^{\text{\acs{LR}}}$ defined above,
the estimated singular channel at frequency $z_k$ is
$\hat{\mathbf H}_{{\rm s},k}
=\sum_j\hat{\mathbf R}_j^{\text{\acs{LR}}}/(z_k-\hat p_j)$.
The regular channel is
$\hat{\mathbf H}_{{\rm reg},k}
=\hat{\mathbf H}_k-\hat{\mathbf H}_{{\rm s},k}$,
where $\hat{\mathbf H}_k=\mathbf H_{{\rm dir},k}+[\hat g_{mn,k}]$ is the measured total \ac{CSI};
$\mathbf H_{{\rm dir},k}=\mathbf H_{\rm dir}(z_k)$ is the known direct field.
Define the singular Gram matrix as
$\mathbf M_{{\rm s},k}
=\hat{\mathbf H}_{{\rm s},k}^{H}\hat{\mathbf H}_{{\rm s},k}
+\xi_k\mathbf I_{N_t}$, where $\xi_k\geq0$ is a numerical ridge.

The frequency-weighted singular-channel exposure is
\begin{equation}
J_{\rm s}(\{\mathbf Q_k\})
\triangleq
\sum_{k=1}^{K}
\beta_k
\operatorname{tr}
\left(
\mathbf M_{{\rm s},k}
\mathbf Q_k
\right),
\label{eq:singular_regular_metrics}
\end{equation}
where $\beta_k\geq0$, $\sum_k\beta_k=1$, are frequency weights and
$\operatorname{tr}(\cdot)$ denotes the matrix trace.
Apart from the numerical ridge,
$\operatorname{tr}(\mathbf M_{{\rm s},k}\mathbf Q_k)$ is the expected
output energy contributed by the estimated singular component at
frequency $k$.

Using the measured total \ac{CSI} $\hat{\mathbf H}_k$, the predicted achievable rate under $\{\mathbf Q_k\}$ is
\begin{equation}
\hat I(\{\mathbf Q_k\})
=
\sum_{k=1}^{K}
\log_2
\det
\left(
\mathbf I_{N_r}
+
\frac{1}{N_0}
\hat{\mathbf H}_k
\mathbf Q_k
\hat{\mathbf H}_k^H
\right),
\label{eq:achievable_rate}
\end{equation}
Let $\hat C_{\text{\acs{WF}}}$ denote the water-filling rate computed using
the singular values of $\hat{\mathbf H}_k$ under the same total
power budget $P$, and let $\delta_C\in[0,1)$ denote the maximum
allowable relative rate loss.

The proposed singularity-aware power allocation is formulated as
\begin{equation}
\begin{aligned}
\min_{\{\mathbf Q_k\succeq0\}}
\quad &
J_{\rm s}(\{\mathbf Q_k\})
\\
\mathrm{s.t.}\quad
&
\hat I(\{\mathbf Q_k\})
\geq
(1-\delta_C)\hat C_{\text{\acs{WF}}},
\\
&
\sum_{k=1}^{K}
\operatorname{tr}
\left(
\mathbf Q_k
\right)
\leq
P,
\end{aligned}
\label{eq:singularity_aware_optimization}
\end{equation}
The rate constraint limits the rate loss relative to water-filling, while
the power cap limits total transmitted power. Since $J_{\rm s}$ is linear
in $\{\mathbf Q_k\}$ and the rate in \eqref{eq:achievable_rate} is
concave in $\{\mathbf Q_k\}$, problem
\eqref{eq:singularity_aware_optimization}
is convex. After obtaining the optimal covariance
$\mathbf Q_k^\star$, its eigendecomposition
\begin{equation}
\mathbf Q_k^\star
=
\mathbf F_k
\boldsymbol\Lambda_k
\mathbf F_k^H
\label{eq:optimal_covariance_eig}
\end{equation}
provides the spatial precoding directions through the unitary matrix
$\mathbf F_k$ and the corresponding power allocation through the
nonnegative diagonal entries of $\boldsymbol\Lambda_k$.

\section{Simulation Results and Analysis}
\subsection{Simulation Setup}
\begin{figure}[t]
    \centering
    \includegraphics[width=0.75\columnwidth]
    {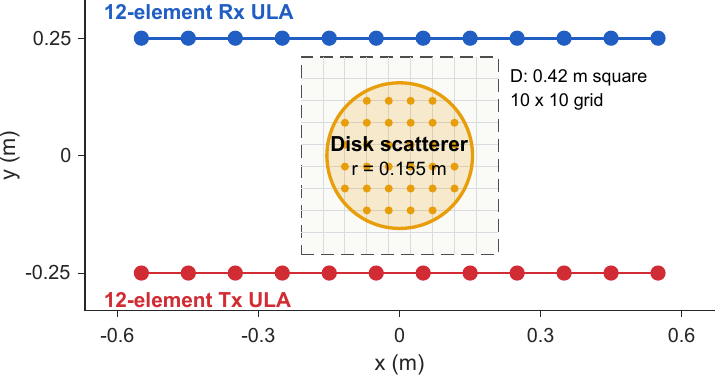}
    \caption{Simulation scenario and \ac{MIMO} array configuration.}
    \label{fig:simulation_scenario}
\end{figure}

The two-dimensional \ac{TM}$_z$ simulation geometry is shown in
Fig.~\ref{fig:simulation_scenario}. A disk of radius $0.155$\,m is placed
in a $0.42$\,m square domain discretized on a $10\times10$ grid, with 32
cells inside the disk. The transmitter and receiver are 12-element
\acp{ULA}
spanning $x\in[-0.55,0.55]$\,m and located at $y=-0.25$\,m and
$y=+0.25$\,m, respectively. The free-space wave speed is
$c_0=3\times10^8$\,m/s. The material contrast is
$\chi_\tau(\mathbf r;z)=\alpha_\tau\chi_0(\mathbf r;z)$, where the
baseline profile $\chi_0$ has unit amplitude and the nominal scale is
$\bar\alpha=1.75$.

Across eight \ac{MIMO} calibration snapshots $\tau=1,\ldots,8$, the scale follows
$\alpha_\tau=\bar\alpha+\rho_\alpha(\alpha_{\tau-1}-\bar\alpha)+w_\tau$
for $\tau\geq2$, with stationary initialization,
$\rho_\alpha=0.78$, $w_\tau\sim\mathcal N(0,0.030^2)$,
$\alpha_\tau\in[1.55,1.95]$ after clipping, and random seed 23.

The \ac{MIMO} band spans 5--10\,GHz. Starting from 21 uniformly spaced tones,
adaptive sampling adds frequencies according to scattered-\ac{CSI} energy and
spectral variation with weights 0.55 and 0.45, respectively, resulting
in 37 nonuniform tones. The communication \ac{SNR} is
15\,dB.
Shared-pole \ac{VF} uses five poles, eight relocation iterations, 12
compressed responses, and up to 100 pole--residue refinement iterations. Low-rank residue recovery retains 98.5\% of the
residue energy with a maximum rank of eight; $\xi_k$ is $10^{-10}$ times
the mean diagonal singular Gram energy. All 37 measured frequencies are used for the main reconstruction test;
no held-out-frequency metric is reported.

For numerical fitting, we use the equivalent GHz coordinate
$\tilde z=z/(2\pi\cdot10^9)$ and pole positions
$\tilde p_j=p_j/(2\pi\cdot10^9)$; the rational form is unchanged after
rescaling its residues. Plotted pole frequencies are $\Re\tilde p_j$;
in figure labels, $\mathbf H(f)$ abbreviates $\mathbf H(2\pi f)$.

The normalized transmit-power budget is $P=1$. Both allocation schemes
use all 12 transmit modes and the same fixed array geometry. The
CVX/SDPT3 solver minimizes the exposure $J_{\rm s}$ subject to a rate
floor of $95\%$ of the full-space \ac{WF} rate and the total-power cap.
The weights $\beta_k$ are normalized trapezoidal weights. Allocation uses
noise-free calibration \ac{CSI}; the communication noise corresponds to
$15$ dB \ac{SNR}.

The reconstruction reference is generated by a discretized
Lippmann--Schwinger forward solver using the \ac{MoM}, rather than by
prescribing poles and residues. The known direct field is subtracted
before fitting the scattered channel. At the final calibration snapshot,
the same complex Gaussian noise pattern is scaled to measurement
\acp{SNR} of $0$, $10$, $20$, and $30$ dB, relative to scattered-channel
energy. The five-pole inverse model has a first-order analytic background;
it is an approximation to the forward electromagnetic solution.

An independent fixed-pair \ac{SISO} experiment is
also conducted with 81
uniformly sampled frequencies and 240 slow-time samples.
This experiment uses a three-pole extraction model.
Its material parameter follows a first-order autoregressive (AR(1)) process with correlation coefficient 0.985 and innovation standard deviation 0.006.
The pole-associated \ac{SISO} component $h_{\rm s}(f;\alpha)$ is extracted after removing direct propagation and fitting residues at the shared poles.
This experiment is independent of the \ac{MIMO} power optimization.

\subsection{Results and Analysis}

\begin{figure}[t]
    \centering
    \includegraphics[width=1.1\columnwidth]{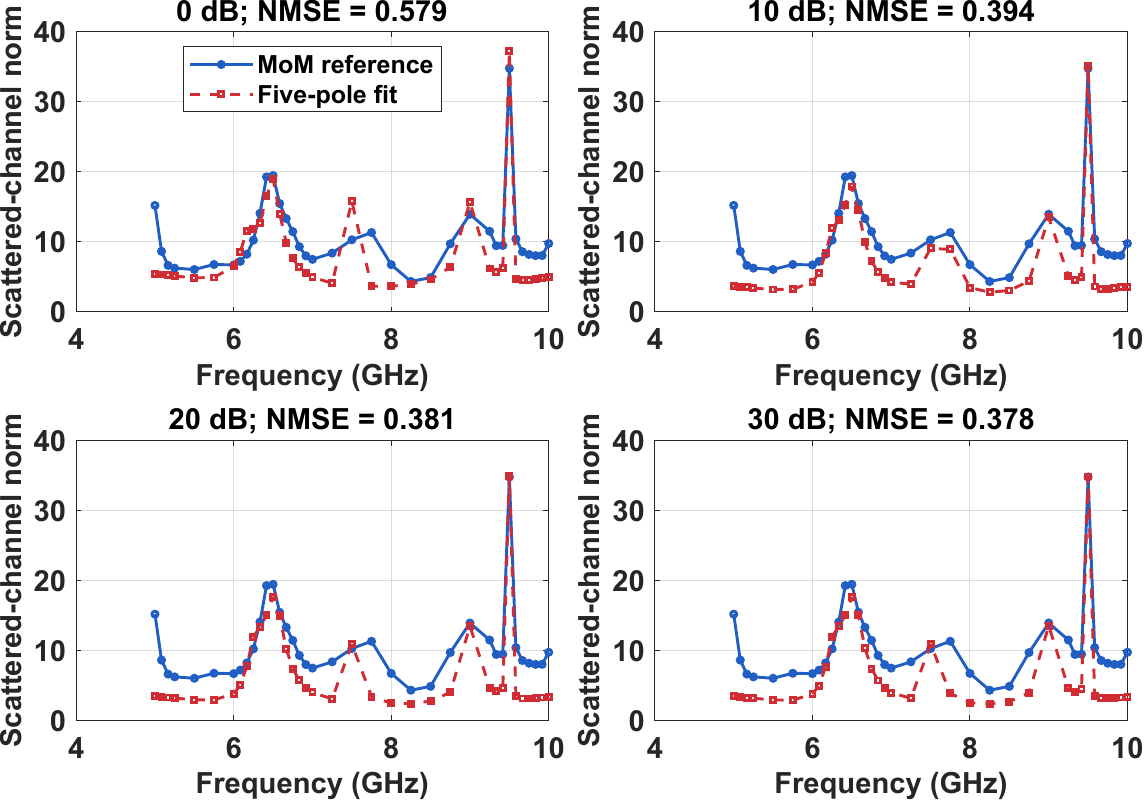}
    \caption{Five-pole reconstruction of independently computed
    scattered electromagnetic \ac{MIMO} channels at four measurement
    \acp{SNR}. The \ac{NMSE} uses all complex matrix entries.}
    \label{fig:singular_reconstruction}
\end{figure}

We first evaluate channel reconstruction against an independently computed electromagnetic reference. 
Fig.~\ref{fig:singular_reconstruction} compares the scattered-channel Frobenius norms from the \ac{MoM} solver and the five-pole inverse model.
Accuracy is assessed on the complete complex channel matrices, rather than their plotted norms alone:
\begin{equation}
\text{\acs{NMSE}}=\frac{\sum_{k=1}^{K}
\|\mathbf H_{{\rm sc},k}^{\rm fit}-\mathbf H_{{\rm sc},k}^{\rm EM}\|_F^2}
{\sum_{k=1}^{K}\|\mathbf H_{{\rm sc},k}^{\rm EM}\|_F^2}.
\label{eq:em_reconstruction_nmse}
\end{equation}
Here $\mathbf H_{{\rm sc},k}^{\rm EM}$ is the noise-free forward solution after direct-field subtraction and $\mathbf H_{{\rm sc},k}^{\rm fit}$ is the reconstructed response at the same measured frequencies.
The full-band \acp{NMSE} at $0$, $10$, $20$, and $30$ dB are $0.579$, $0.394$, $0.381$, $0.378$, respectively.
The noise-free fit gives an \ac{NMSE} of $0.376$.
Thus, increasing measurement \ac{SNR} reduces the noise contribution, but the error saturates because five poles and a linear background do not fully represent the electromagnetic frequency response.
These results distinguish full-response error from recovery of singular frequency structure.
Relative to the noise-free electromagnetic fit, the maximum shift of the five reconstructed pole frequencies is $23.85$ MHz at $20$ dB and $13.04$ MHz at $30$ dB, below the minimum sample spacing of $83.33$ MHz.
The dominant pole near $9.5$ GHz shifts by only $0.18$ MHz at $30$ dB.
Thus, appreciable matrix-fit error can coexist with stable identification of the dominant singular frequencies.
Pole locations are the key singularity information: they determine the resonant denominators and the $|z-p_j|^{-2}$ sensitivity in \eqref{eq:pole_channel_perturbation}, whereas residues govern coupling strength and spatial directions.
These comparisons establish numerical pole stability, not absolute physical-pole accuracy, since the forward solver supplies no independent complex-pole labels.

\begin{figure}[t]
    \centering
    \includegraphics[width=0.8\columnwidth]
    {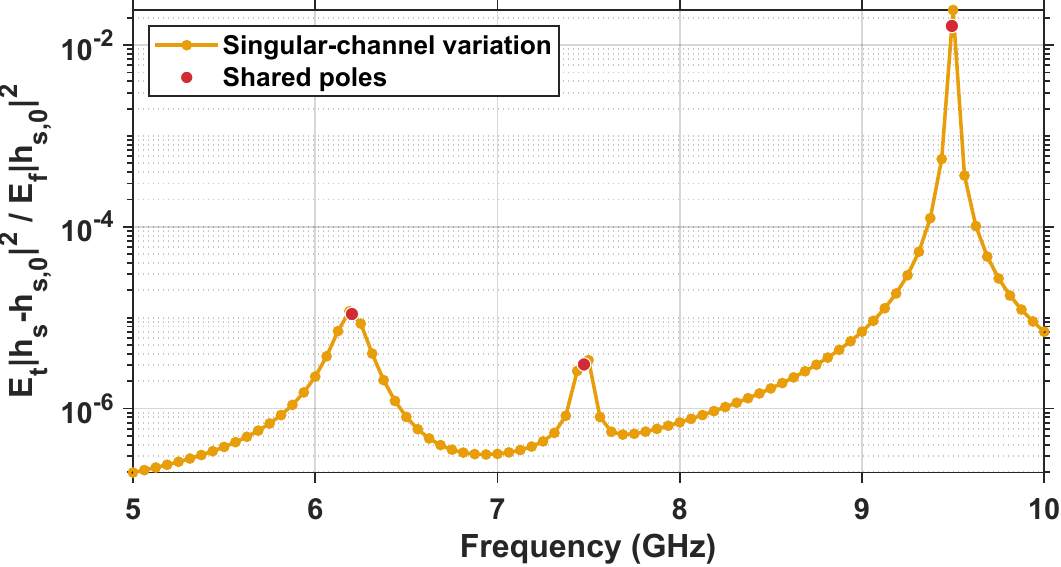}
    \caption{Normalized \ac{SISO} singular-channel variation versus frequency
    under the same environmental trajectory.}
    \label{fig:normalized_csi_variation}
\end{figure}

Fig.~\ref{fig:normalized_csi_variation} shows the \ac{SISO} singular-component variation $\mathbb E_\tau|h_{\rm s}(f;\alpha_\tau)-h_{\rm s}(f;\alpha_1)|^2/ \langle|h_{\rm s}(f;\alpha_1)|^2\rangle_f$, where $\langle\cdot\rangle_f$ denotes the 81-tone average.
Clear peaks appear near the shared poles, where the normalized variations reach $1.10\times10^{-5}$, $3.06\times10^{-6}$, and $1.63\times10^{-2}$, respectively.
Under the same material-variation trajectory, the singular component is more sensitive to environmental perturbations near singular frequencies.
This agrees with the preceding analysis that pole-associated singularities amplify channel perturbations and reduce channel stability.

\begin{figure}[t]
    \centering
    \hspace{-0.12\columnwidth}
    \includegraphics[width=0.8\columnwidth]{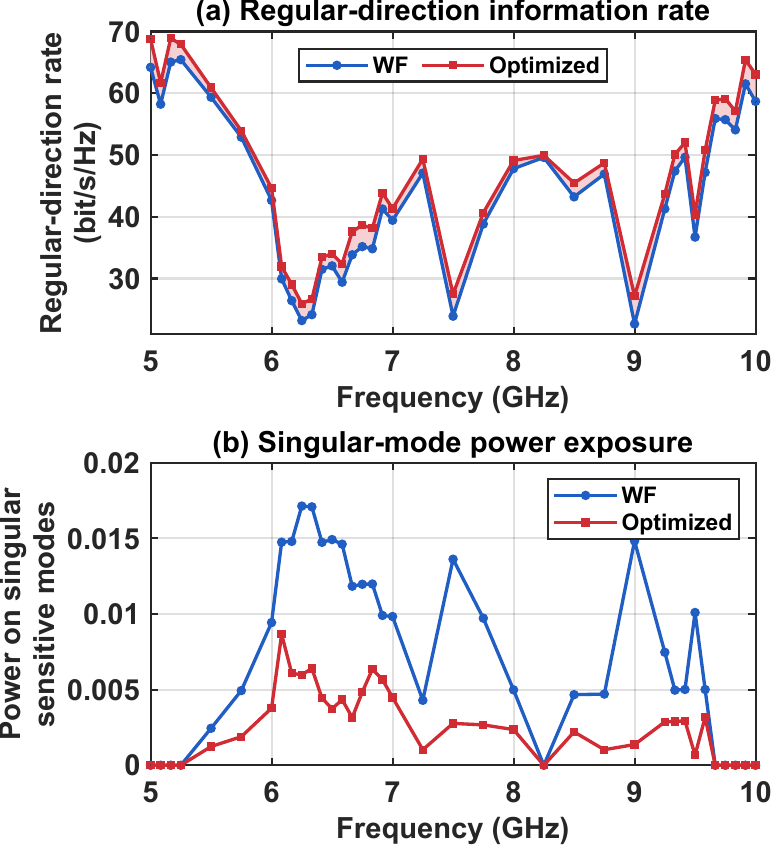}
    \caption{Power redistribution under the same fixed array:
    (a) regular-direction information rate and
    (b) power allocated to singular-sensitive modes.}
    \label{fig:modal_fragility}
\end{figure}

Fig.~\ref{fig:modal_fragility} examines the power-redistribution mechanism
using the saved \ac{WF} and optimized covariances from this fixed-array
electromagnetic experiment. Diagonalizing
$\hat{\mathbf H}_{{\rm sing},k}^{H}
\hat{\mathbf H}_{{\rm sing},k}$ gives a common transmit basis for both
schemes. Modes above the global upper-quartile eigenvalue threshold are
classified as singular-sensitive; the remaining modes form
$\mathbf V_{{\rm reg},k}$ and the projector
$\mathbf P_{{\rm reg},k}=\mathbf V_{{\rm reg},k}\mathbf V_{{\rm reg},k}^{H}$.
For $a\in\{\text{\acs{WF}},\text{\acs{SA}}\}$, define
\begin{equation}
 \mathbf Q_{{\rm reg},k}^{a}
 =\mathbf P_{{\rm reg},k}\mathbf Q_k^{a}\mathbf P_{{\rm reg},k}.
\end{equation}
Fig.~\ref{fig:modal_fragility} (a) uses the per-tone log-determinant in \eqref{eq:achievable_rate}
with $\widehat{\mathbf H}_{{\rm reg},k}$ and
$\mathbf Q_{{\rm reg},k}^{a}$. 
The regular component comprises the regular channel part defined in Section III-B.
Neither power nor noise is rescaled.
Aggregate regular-direction power increases from $0.73$ to $0.90$; the mean information rate increases from $43.70$ to $46.43$ bit/s/Hz ($6.25\%$), with gains at all 37 tones.

Fig.~\ref{fig:modal_fragility} (b) plots $\operatorname{tr}[(\mathbf I-\mathbf P_{{\rm reg},k})
\mathbf Q_k^{a}]$. Its frequency sum decreases from $0.27$ to $0.10$
($64.03\%$), showing reduced power in singular-sensitive directions.
This unweighted modal-power metric differs from $J_{\rm s}$, which
weights power by singular-channel energy and frequency: $J_{\rm s}$
decreases from $0.17$ to $0.02$ ($90.32\%$). Meanwhile, the total-channel
nominal rate changes from $70.75$ to $67.21$ bit/s/Hz per tone,
retaining $95\%$ of the \ac{WF} rate. 
Fig.~\ref{fig:modal_fragility} (a) shows a rate gain for the regular-channel component under the projected covariances, not an increase in total-channel capacity.
This evaluation does not imply an additive decomposition of that capacity.

\begin{figure}[!t]
    \centering
    \includegraphics[width=0.96\columnwidth]
    {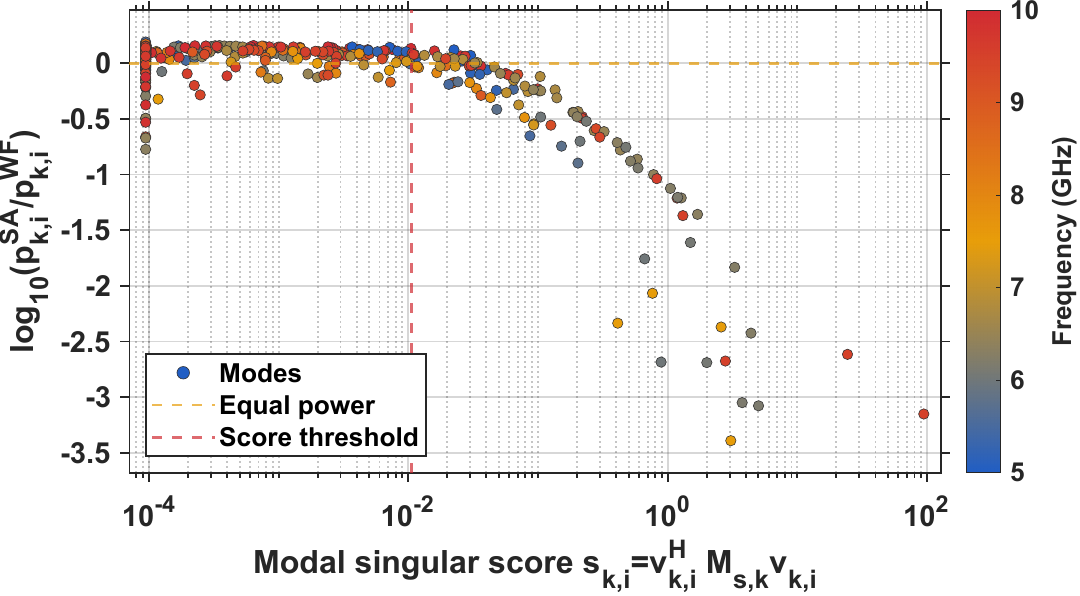}
    \caption{Relationship between the singularity score and the modal
    power change relative to conventional \ac{WF}.}
    \label{fig:fragility_power_correlation}
\end{figure}

The power-redistribution mechanism is examined using the modal singular
score $s_{k,i}=\mathbf v_{k,i}^{H}\mathbf M_{{\rm s},k}\mathbf v_{k,i}$,
where $\mathbf v_{k,i}$ is a nominal right singular vector of the total
channel. 
This score characterizes coupling to the fitted singular component.
Fig.~\ref{fig:fragility_power_correlation} provides a complementary illustration of this mechanism using a separate power-allocation experiment with a three-pole model.

Fig.~\ref{fig:fragility_power_correlation} relates the same score to
the relative power change
$
\Delta p_{k,i}
=
\log_{10}
\left(
p_{k,i}^{\text{\acs{SA}}}/p_{k,i}^{\text{\acs{WF}}}
\right).
$
Here $p_{k,i}^{m}=\mathbf v_{k,i}^H\mathbf Q_k^{m}\mathbf v_{k,i}$
for $m\in\{\text{\acs{SA}},\text{\acs{WF}}\}$, with
$\mathbf Q_k^{\text{\acs{SA}}}=\mathbf Q_k^\star$; the plot includes nominal
\ac{WF}-active modes.
The horizontal dashed line $\Delta p_{k,i}=0$ indicates unchanged power
relative to \ac{WF};
points below it correspond to power suppression.

Low-score modes cluster near $\Delta p_{k,i}\approx0$ and thus largely
retain their \ac{WF} power. As $s_{k,i}$ increases, the points move downward:
$\Delta p_{k,i}=-1$, $-2$, and $-3$ correspond to $10\%$, $1\%$, and
$0.1\%$ of \ac{WF} power, and several high-score modes reach $-2$ to $-3.5$.
The red dashed line marks the upper-quartile score threshold; modes to
its right generally receive larger power reductions. Scatter reflects
frequency weights, channel gains, available modes, and the rate floor.
The correlation of $-0.404$ thus indicates a moderate negative trend,
not a one-to-one relation. Colors show that high-score modes occur at
multiple frequencies. The method therefore reduces singular-channel
coupling through frequency--spatial power redistribution while retaining
most of the conventional \ac{WF} transmission capability.

\section{Conclusion}

This paper derived an electromagnetic \ac{MIMO} model that separates regular
and pole-associated channel components through the scattering resolvent,
linking electromagnetic poles to environmental sensitivity. Shared poles
are estimated from broadband \ac{CSI}, and transmit covariance is optimized
under a \ac{WF}-relative rate floor. Independent electromagnetic data expose the finite-order reconstruction
error, while the separate slow-time experiment illustrates pole-associated
sensitivity. The allocation reduces weighted singular
exposure by approximately $90.32\%$ while retaining $95\%$ of the nominal \ac{WF} rate.


\end{document}